\documentclass[aps,prl,showpacs,superscriptaddress,twocolumn]{revtex4-1}
\usepackage{graphicx}
\usepackage{amssymb}
\usepackage{dcolumn}
\usepackage{bm} 
\begin{document}
\title{Fermi surface and superconductivity in low-density high-mobility $\delta$-doped SrTiO$_{3}$}
\author{M. Kim}
\affiliation{Department of Advanced Materials Science, University of Tokyo, Kashiwa, Chiba 277-8561, Japan}
\author{C. Bell}
\affiliation{Department of Advanced Materials Science, University of Tokyo, Kashiwa, Chiba 277-8561, Japan}
\affiliation{Department of Applied Physics and Stanford Institute for Materials and Energy Science, Stanford University, Stanford, California 94305, USA}
\author{Y. Kozuka}
\affiliation{Department of Advanced Materials Science, University of Tokyo, Kashiwa, Chiba 277-8561, Japan}
\author{M. Kurita}
\affiliation{Department of Advanced Materials Science, University of Tokyo, Kashiwa, Chiba 277-8561, Japan}
\author{Y. Hikita}
\affiliation{Department of Advanced Materials Science, University of Tokyo, Kashiwa, Chiba 277-8561, Japan}
\author{H. Y. Hwang}
\affiliation{Department of Advanced Materials Science, University of Tokyo, Kashiwa, Chiba 277-8561, Japan}
\affiliation{Department of Applied Physics and Stanford Institute for Materials and Energy Science, Stanford University, Stanford, California 94305, USA}
\affiliation{Japan Science and Technology Agency, Kawaguchi, Saitama 332-0012, Japan}

\begin{abstract}
The electronic structure of low-density $n$-type SrTiO$_{3}$ $\delta$-doped heterostructures is investigated by angular dependent Shubnikov-de Haas oscillations. In addition to a controllable crossover from a three- to two-dimensional Fermi surface, clear beating patterns for decreasing dopant layer thicknesses are found. These indicate the lifting of the degeneracy of the conduction band due to subband quantization in the two-dimensional limit. Analysis of the temperature-dependent oscillations shows that similar effective masses are found for all components, associated with the splitting of the light electron pocket. The dimensionality crossover in the superconducting state is found to be distinct from the normal state, resulting in a rich phase diagram as a function of dopant layer thickness.
\end{abstract}

\maketitle
	
	 The burgeoning field of oxide electronics is driven by the rich variety of their physical properties, suggesting great potential for future multifunctional devices \cite{takagi_science2010,Mannhart_Science2010}. A workhorse material in this field, SrTiO$_{3}$ (STO), is widely used as a perovskite substrate. However, STO itself displays a range of intriguing low temperature phenomena. For example, it is a quantum paraelectric, showing very high permittivity \cite{Muller_PRB1979}, enabling effective screening of impurities. Thus $n$-doped STO is a high-mobility, low-density superconducting semiconductor \cite{Tufte_PR1967,Frederikse_PR1967,schooley_prl1964,Hulm_PLTP1970}.\par
	 
	Recently, these aspects have motivated the exploration of two-dimensional (2D) electron physics in STO, leading to a variety of heterostructure implementations, notably the LaAlO$_{3}$/SrTiO$_{3}$ (LAO/STO) interface \cite{Ohtomo_nature2004}, (ferroelectric) field effect transistors \cite{Takahashi_Nature2006,ueno_natmat2008,Lee_PRL2011}, and $\delta$-doped systems \cite{Kozuka_Nat2009}. In the latter system, 2D superconductivity (SC) and 2D Shubnikov-de Haas (SdH) quantum oscillations were simultaneously observed. The LAO/STO interface also displays a similar fascinating combination of properties \cite{reyren_science2007,Caviglia_PRL2010_2,BenShalom_PRL2010,Huijben_condmat2010}. Angle-resolved photoemission spectroscopy (ARPES) has even shown that 2D electron states can be induced on cleaved STO surfaces \cite{Santander-Syro_Nature2011,Meevasana_NatMater2011}.  

	Bulk STO is a $d^{0}$-electron system with a three-fold degenerate conduction band at the $\Gamma$ point, formed by the $d_{\mathrm{xy}}$, $d_{\mathrm{yz}}$, and $d_{\mathrm{zx}}$ $t_{\mathrm{2g}}$ orbitals. The added effects of a tetragonal distortion at $\sim$ 105 K and the spin-orbit interaction mix these orbital characteristics, depending on their energy scale relative to the Fermi energy, as described by first-principles studies \cite{mattheis_PRB1972b,Bistritzer_PRB2011}. Due to these complexities, many questions remain, particularly in relation to the symmetry, effective mass and electronic structure of these 2D STO systems. This is in part due to the complex form and small amplitude of the SdH oscillations \cite{Kozuka_Nat2009,Caviglia_PRL2010_2,BenShalom_PRL2010,Jalan_PRB2010}, which hinders analysis.\par
	
	To further our understanding, here we use $\delta$-doping to realize 2D electron states in STO, avoiding lattice mismatch and interface or surface scattering. This technique has been developed using pulsed laser deposition \cite{Kozuka_Nat2009, Kozuka_APL2010_1,Kozuka_APL2010_2}, as used here, as well as molecular beam epitaxy \cite{Jalan_PRB2010}. We deposited STO heterostructures, 0.2 at$.$ \% Nb:SrTiO$_{3}$ (NSTO) with varying thickness $d_{\mathrm{NSTO}}$ in the range 11 nm $ \le d_{\mathrm{NSTO}} \le$ 292 nm sandwiched by 100 nm thick undoped STO cap and buffer layers, on STO (001) substrates. Details of the growth technique and studies using 1 at$.$ \% NSTO have been reported previously \cite{Kozuka_APL2010_1,Kozuka_APL2010_2}. We stress several points: we could achieve bulk-quality NSTO films for this level of doping \cite{Tufte_PR1967,Frederikse_PR1967}, and a further enhancement of the mobility up to 4900 cm$^{2}$/Vs by $\delta$-doping, resulting in a dramatic increase in the SdH oscillations compared to the 1 at$.$ \% NSTO samples.
		
	\begin{figure}[bp]
	\begin{center}
	\includegraphics{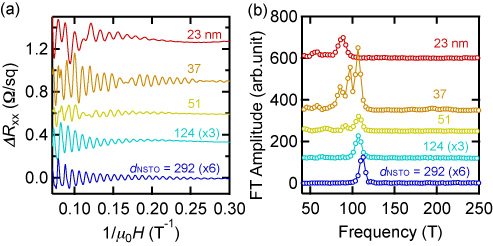}
	\end{center}
	\caption{(color online) (a) Shubnikov-de Haas (SdH) oscillations with varying $d_{\mathrm{NSTO}}$; $T =$ 100 mK. (b) Fourier transformation (FT) data of (a). Data sets in (a) and (b) have been offset vertically and scaled for clarity.}
	\label{fig1}
	\end{figure}
	
	Magneto-transport measurements at a temperature of $T = 100$ mK revealed clear SdH oscillations as shown in Fig$.$$\ \ref{fig1}$ (a). The oscillatory components here were extracted by fitting the positive magnetoresistance background using a polynomial function. It is notable that clear beating patterns were observed only for relatively small $d_{\mathrm{NSTO}}$. Fourier transforms (FT) of the data, shown Fig$.$$\ \ref{fig1}$ (b), show the appearance of peak splitting in the spectra as $d_{\mathrm{NSTO}}$ decreases: for example, three frequencies are clear for $d_{\mathrm{NSTO}} = $ 37 nm. Immediately we can infer that since the beat patterns only appear in the thinner samples, these multifrequencies are associated with the lifting of the degenerate bulk bands by 2D subband quantization in the substrate normal direction, similar to calculations for STO-based heterostructures \cite{Popovic_PRL2008}.\par

\begin{figure}[tbp]
	\begin{center}
	\includegraphics{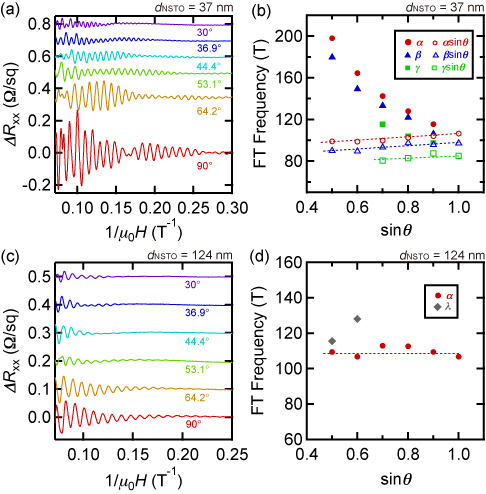}
	\end{center}
	\caption{(color online) SdH oscillations at $T = 100$ mK, for various $\theta$, and the positions of the peaks in the FT spectra, vs. sin$\theta$, for two samples: (a) and (b), $d_{\mathrm{NSTO}}$ = 37 nm, (c) and (d) $d_{\mathrm{NSTO}}$ = 124 nm. Solid symbols in (b) and (d) show the raw FT frequencies, while open symbols in (b) are scaled by sin$\theta$. Dotted lines are guides to the eye.}
	\label{fig2}
	\end{figure}

	To confirm the dimensionality of the samples, we measured the SdH oscillations for various magnetic field orientations, where the angle $\theta = 90^{\circ}$ corresponds to the magnetic field $H$ perpendicular to the dopant plane. We focus on two representative samples, with $d_{\mathrm{NSTO}}$ = 37 nm and 124 nm. Although for $d_{\mathrm{NSTO}}$ = 37 nm, the shape of the SdH oscillations changes in a complicated manner [Fig$.$$\ \ref{fig2}$ (a)], the peak frequencies extracted from the FT analyses show a clear trend as a function of sin$\theta$, as plotted in Fig$.$$\ \ref{fig2}$ (b) (solid symbols). Here we label each peak as $\alpha$, $\beta$ and $\gamma$ in descending order from the highest frequency. If we multiply the peak frequencies by sin$\theta$ [open symbols in Fig$.$$\ \ref{fig2}$ (b)], the resultant data are almost independent of $\theta$. This proportionality to the perpendicular magnetic field component confirms the 2D character of the SdH oscillations. The slight decrease of the peak frequencies can be understood by magnetic freeze-out effects.\par

       In contrast, for the $d_{\mathrm{NSTO}}$ = 124 nm sample, while the amplitude of the signal decreases as $\theta$ decreases [Fig$.$$\ \ref{fig2}$(c)], the oscillation frequency is essentially unchanged as confirmed in Fig$.$$\ \ref{fig2}$ (d). This indicates a bulk-like 3D Fermi surface as expected when quantization effects are not important. The appearance of an additional peak in the FT spectra, denoted as $\lambda$ in Fig$.$$\ \ref{fig2}$ (d), can be explained by magnetic breakdown around the Fermi surface, as observed previously \cite{Uwe_JJAP1985b}. In the analysis that follows, for convenience we refer to the 37 nm and 124 nm samples as the 2D and 3D samples, respectively.\par

	\begin{figure*}[tb]
	\begin{center}
	\includegraphics{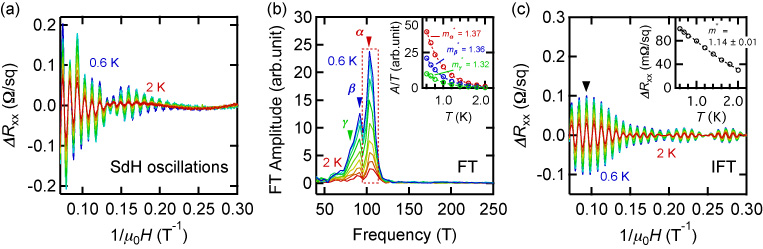}
	\end{center}
	\caption{(a) SdH oscillations for $d_{\mathrm{NSTO}}$ = 37 nm (2D) at various $T$ in the range 0.6 K $\le T \le $ 2.0 K. (b) FT of data in (a). Arrows indicate the different peaks labeled $\alpha$ (red), $\beta$ (blue), and $\gamma$ (green), respectively. Inset: amplitude variation of each peak vs $T$. Dotted lines are fits to Eq$.$$\ \ref{eqS1}$. (c) IFT result for the $\alpha$ peak. Frequency window is indicated by the dotted box in (b). Inset: amplitude variation with $T$ of the peak indicated by the black arrow in the main graph. Dotted line is the theoretical fit.}
	\label{fig3}
	\end{figure*}

	Next we focus on the effective mass $m^{*}$, which we can estimate from the temperature dependence of the SdH oscillation amplitude. For the 2D sample, the SdH data measured at various $T$ are shown in Fig$.$$\ \ref{fig3}$ (a). The temperature evolution of the peak intensities from the FT spectra is shown in Fig$.$$\ \ref{fig3}$ (b), which we fit using the Lifshitz-Kosevich (LF) formula,
	\begin{equation}
	A(T)/T=A_{\mathrm{0}}/\sinh X, \label{eqS1}
	\end{equation}
	where $A_{\mathrm{0}}$ is a prefactor independent of $T$, and $X=14.69m^{\mathrm{*}}T/\mu_{\mathrm{0}}H$. As shown in the inset of Fig$.$$\ \ref{fig3}$ (b), using $\mu_{\mathrm{0}}H$ = 8.5 $\pm$ 5.5 T the estimated effective mass of each peak is $m_{\mathrm{\alpha}}^{\mathrm{*}}$ = 1.37$m^{\mathrm{*}}$, $m_{\mathrm{\beta}}^{\mathrm{*}}$ = 1.36$m^{\mathrm{*}}$, and $m_{\mathrm{\gamma}}^{\mathrm{*}}$ = 1.32$m^{\mathrm{*}}$, respectively. These similar values indicate that these electrons may be split from same electron pocket, supporting the 2D subband quantization picture.\par
	
	Due to the relatively large field range used in this analysis, and the correspondingly large error in the values of $m^{\mathrm{*}}$, we performed a cross-check using the inverse FT (IFT) of one of the peaks in the FT spectra. After performing an IFT on the data within the window defined by the red box in Fig$.$$\ \ref{fig3}$ (b), the $T$ dependence of a single peak at $1/\mu_{\mathrm{0}} H \sim$ 0$.$093 T$^{-1}$ was followed, as indicated in Fig$.$$\ \ref{fig3}$ (c). This analysis provides a more accurate $m^{\mathrm{*}}$ via the fit to Eq$.$$\ \ref{eqS1}$, [inset of Fig$.$$\ \ref{fig3}$ (c)], resulting in $m_{\mathrm{\alpha}}^{\mathrm{*}}$ = 1.14 $\pm$ 0.01, in good agreement with the FT analysis. Since the SdH data for the 3D sample were dominated by a single frequency, $m^{\mathrm{*}}$ in this case could be estimated directly, giving $m^{\mathrm{*}}$ = 1.12 $\pm$ 0.02. Both of these values are comparable to other transport studies in STO bulk and heterostructures, but significantly different from ARPES experiments on cleaved samples, where $m^{\mathrm{*}} \sim $ 0$.$6, as summarized in Table \ref{masstable}. The reason for this may be associated with electron-phonon renormalization effects \cite{Meevasana_NJP2010}.

	\begin{table}[b]
	\caption{The effective mass of light electrons estimated for various STO samples and heterostructures.}
	\begin{center}
	\begin{tabular}{c|c|c|c} 
	\hline
	Method & Sample & $m^{*}$ & Reference \\ \hline\hline
	& Bulk NSTO & 1$.$24-1$.$5 & \cite{Uwe_JJAP1985b,Frederikse_PR1967_2} \\
	SdH & 1 \% $\delta$-doped STO & 1$.$24-1$.$56 & \cite{Kozuka_Nat2009,Jalan_PRB2010} \\
	& LAO/STO & 1$.$45-2$.$1 & \cite{Caviglia_PRL2010_2,BenShalom_PRL2010} \\
	& 0.2 \% $\delta$-doped STO & 1$.$12-1$.$38 & This work \\ \hline
	ARPES & Bulk NSTO & 1$.$2-1$.$3 & \cite{chang_prb2010,Meevasana_NJP2010} \\
	& Cleaved STO surface & 0$.$5-0$.$7 & \cite{Santander-Syro_Nature2011,Meevasana_NatMater2011} \\ \hline
	\end{tabular}
	\end{center}
	\label{masstable}
	\end{table}

	\begin{figure}[b]
	\begin{center}
	\includegraphics[width=7.8cm]{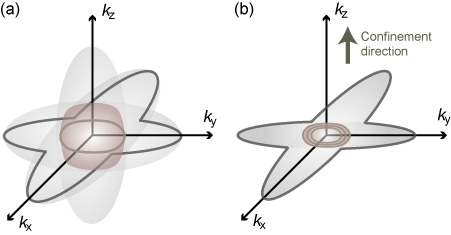}
	\end{center}
	\caption{(color online) Schematic diagram of the STO Fermi surface for (a) $d_{\mathrm{NSTO}}$ = 124 nm (3D) and (b) $d_{\mathrm{NSTO}}$ = 37 nm (2D). The extremum of the Fermi surface in the [001] direction is shown with a thick line for each electron pocket.}
	\label{fig4}
	\end{figure}

	Based on these results we can draw schematic Fermi surfaces for the 3D and 2D samples, as shown in Fig$.$$\ \ref{fig4}$. The two extrema of the Fermi surface in the (001) direction are formed by the inner circular pocket of light electrons, and the outer `star' pocket of heavy electrons. We note that we could not detect SdH oscillations from the heavy electrons in these samples, although their existence is inferred from other measurements. Estimates of the heavy effective mass varies from $m^{\mathrm{*}}$ = 7 \cite{chang_prb2010} to 10 - 20 \cite{Santander-Syro_Nature2011}, thus the resulting low cyclotron frequency, $\omega_{\mathrm{c}} = e\mu_{\mathrm{0}}H/m^{*}$, ($e$ is the electronic charge), makes the detection via SdH oscillations more challenging. We can conclude that in the 3D sample, the observed electron pocket is the inner circular one, while for the 2D sample, it is natural to associate the observed multiple frequencies, all with the same $m^{\mathrm{*}}$, as the same light electrons, but split by 2D subband quantization.
	
	The existence of non-oscillating heavy electrons sheds light on the observed discrepancy between the sheet carrier density estimated from Hall measurements, $N_{\mathrm{Hall}}$ (at $\mu_{\mathrm{0}}H = 1$ T), and the SdH oscillation density, $N_{\mathrm{SdH}}$ via the Onsager relation. Such missing carriers have been observed previously \cite{Kozuka_Nat2009,Caviglia_PRL2010_2,BenShalom_PRL2010,Jalan_PRB2010,Huijben_condmat2010}. The ratio of $N_{\mathrm{Hall}}/N_{\mathrm{SdH}}$ is 3$.$5 in the 3D sample and 5$.$3 in the 2D sample, the slight difference possibly due to elongation of the electron pocket by the confining potential \cite{Santander-Syro_Nature2011}. Notably, the existence of this discrepancy for all $d_{\mathrm{NSTO}}$ implies that this multiband character is intrinsic, explained consistently via band structure considerations, and not dominated by spatial variations of the electron density or mobility. Additionally, in contrast to the asymmetric potential wells formed at the STO surface, or the LAO/STO heterointerface, here our potential is designed to be symmetric, and added complexities such as the presence of Rashba splitting are avoided. 

	Here we assumed spin degeneracy is not lifted in the Onsager relationship. However, if we take an effective electron $g$-factor of $g^* = 2$, we find the unusual case where the Zeeman splitting, $\Delta_{\mathrm{Zeeman}} = g^{*}\mu_{\mathrm{B}}\mu_{\mathrm{0}}H$ is always larger than the Landau splitting $\Delta_{\mathrm{Landau}} = \hbar \omega_{\mathrm{c}}$ ($\mu_{\mathrm{B}}$ is the Bohr magneton, $\hbar$ is the Plank constant divided by 2$\pi$). In this case, spin degeneracy is always lifted. We note that the 37 nm data in Fig$.$$\ \ref{fig1}$ (a) shows signs of peak splitting at the highest magnetic field. If this peak splitting is associated with Zeeman splitting, this would contradict the notion that $\Delta_{\mathrm{Zeeman}}  > \Delta_{\mathrm{Landau}} $. The more conventional case ($\Delta_{\mathrm{Zeeman}}  <  \Delta_{\mathrm{Landau}} $) would then  imply that $g^* < 2$. Alternatively this splitting may be due to different sub-bands which are not resolved, due to disorder broadening, at lower fields. Further angular dependent measurements at higher fields may clarify these points \cite{Fang_PR1968}.

\begin{figure}[t]
	\begin{center}
	\includegraphics{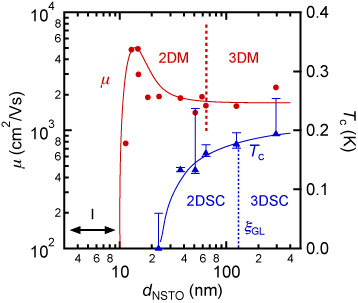}
	\end{center}
	\caption{(color online) Low-temperature Hall mobility $\mu$ and superconducting transition temperature $T_\mathrm{{c}}$ vs. $d_{\mathrm{NSTO}}$. Error bars for $T_\mathrm{{c}}$ denote the 10\%-90\% width of the superconducting transition. Regions are denoted I: insulating, 2DM: 2D metallic, 3DM: 3D metallic, 2DSC: 2D superconducting, and 3DSC: 3D superconducting. Solid lines are guides to the eye, dotted lines show estimates of the 3D-2D dimensional crossover for the normal and superconducting states.}
	\label{fig5}
	\end{figure}

	These samples are also superconducting, with mean field critical temperatures ($T_\mathrm{{c}}$) in the range $T_\mathrm{{c}} \le$ 200 mK. Combining these data with the $T = 2$ K Hall mobility $\mu$, the normal state and superconducting phase space of these 0$.$2 at$.$ \% NSTO samples can be mapped, as shown in Fig$.$$\ \ref{fig5}$. Studies of the superconducting critical fields (to be discussed in detail elsewhere) demonstrated a transition from 3D to 2D superconductivity (2DSC), at $d_{\mathrm{NSTO}}$ $\sim$ 124 nm, consistent with estimates of the Ginzburg-Landau coherence length, as expected for 2DSC \cite{Tinkham_PR1963}. In contrast, the dimensional crossover of the normal state into the 2D limit occurs around $d_{\mathrm{NSTO}}$ = 64 nm. Below $d_{\mathrm{NSTO}}$ = 11 nm, the samples were insulating. Thus for a wide window 23 nm $\le d_{\mathrm{NSTO}} \le$ 64 nm 2DSC and 2D metallic states coexist, in a thickness regime significantly easier to control than for the 1 at$.$ \% NSTO samples studied previously. This provides an ideal system to investigate quantum phase transitions into a variety of ground states \cite{Goldman_PT1998}, and novel superconductivity in the 2D clean limit \cite{Rasolt_RMP1992,Casalbuoni_RMP2004}.\par
	
	The high resolution data presented in here reveal the electronic structure of NSTO, and suggest the possibility of further high mobility or mesoscopic effects, such as the quantum Hall effect (QHE). Of particular interest is the possibility of correlations from the $d$ electrons leading to novel physics. It should be noted that the complex band structure of STO will in principle not disrupt the QHE, as demonstrated by $p$-type GaAs \cite{Stormer_PRL1983}, which has the same symmetry as the STO conduction band.

\begin{acknowledgments}
The authors thank P$.$ D$.$ C$.$ King, A. Macdonald, and W$.$ Meevasana for discussions. H$.$Y$.$H$.$ and C$.$B$.$ acknowledge support by the Department of Energy, Office of Basic Energy Sciences, Division of Materials Sciences and Engineering, under contract DE-AC02-76SF00515. M$.$K$.$ acknowledges support from the Japanese Government Scholarship Program of the Ministry of Education, Culture, Sport, Science and Technology, Japan.
\end{acknowledgments}

\end{document}